\documentclass{osa-article}

\journal{osajournal}

\articletype{Research Article}
\usepackage{soul}
\usepackage{lineno}
\usepackage{siunitx}

\begin{document}

\title{Formation of Quasi-Bound States in the Continuum in a Single Deformed Microcavity}

\author{Shuai Liu,\authormark{1,*} Bo-Han Wu,\authormark{2} Jeffrey Huang,\authormark{1} and Zheshen Zhang\authormark{1,*}}

\address{\authormark{1} Department of Electrical Engineering and Computer Science, The University of Michigan, Ann Arbor, Michigan 48109, USA.\\
\authormark{2}Research Laboratory of Electronics, Massachusetts Institute of Technology, Cambridge, Massachusetts, 02139, USA\\
}

\email{\authormark{*}shualiu@umich.edu\\ \authormark{*}zszh@umich.edu} 



\begin{abstract}

Bound states in the continuum (BIC) holds significant promise in manipulating electromagnetic fields and reducing losses in optical structures, leading to advancements in both fundamental research and practical applications. Despite their observation in various optical systems, the behavior of BIC in whispering-gallery-modes (WGMs) optical microcavities, essential components of photonic integrated chips, has yet to be thoroughly explored. In this study, we propose and experimentally identify a robust mechanism for generating quasi-BIC in a single deformed microcavity. By introducing boundary deformations, we construct stable unidirectional radiation channels as leaking continuum shared by different resonant modes and experimentally verify their external strong mode coupling. This results in drastically suppressed leaking loss of one originally long-lived resonance, manifested as more than a 3-fold enhancement of its quality (\textit{Q}) factor, while the other short-lived resonance becomes more lossy, demonstrating the formation of Friedrich-Wintgen quasi-BICs as corroborated by both the theoretical model and the experimental data. This research will provide a practical approach to enhance the \textit{Q} factor of optical microcavities, opening up potential applications in the area of deformed microcavities, nonlinear optics, quantum optics, and integrated photonics.
\end{abstract}

\section{Introduction}

Bound states in the continuum (BIC) was initially proposed nearly a century ago by Neumann and Wigner in a quantum mechanical system \cite{Neumann and Wiger}. Despite residing within the continuous spectrum, these states exhibit perfect localization without any radiation, attributed to destructive interference of waves \cite{NRS ADStone}. In recent years, this counterintuitive mechanism has been discovered across various domains, including acoustic, water, quantum wire, and electromagnetic wave systems \cite{BIC Acoustic, BIC Water, BIC Quantum Wire, BIC Electrics}. Since 2008, BIC has been revisited in coupled optical waveguides structures \cite{PRL 1st Optical BIC}, sparking growing interest for photonic applications. In an ideal setting, BIC can facilitate extremely compact mode confinement and infinite \textit{Q} factors, leading to significantly enhanced light-matter interactions. This unique property positions BIC as an innovative platform for localizing electromagnetic waves while mitigating losses in photonic structures. More recently, advancements in micro/nanofabrication technology have enabled the realization of photonic BICs in various optical systems, promising notable improvements in multiple crucial areas, such as ultralow threshold lasers, nonlinear optics, optical sensing, and topological insulators \cite{Nature BIC Laser1, Nature BIC Laser2, Science Nonlinear, PRL Topological, PNAS Sensing1, Science Sensing2}.

The straightforward approach to creating BIC involves introducing a symmetry mismatch into periodic structures, such as photonic crystals, gratings, and metasurfaces \cite{Nanophotonics Review, PRL PC, Nature Xiao, NP Grating}. In these structures, BICs are realized by preventing the coupling between discrete modes and radiation modes, as they belong to symmetry-categorized eigenmodes. However, this advantage can also pose challenges, since symmetry-protected BIC makes it difficult to efficiently collect emission of these ultra-high-\textit{Q} modes due to the forbidden leaking. Therefore, recent research has focused more on quasi-BIC instead by slightly breaking the BIC condition, which can maintain a sufficiently high-\textit{Q} factor while significantly enhancing emission intensity \cite{NRP Review}. Nevertheless, the periodic structures are not always necessary. Zou \emph{et al.} proposed a novel design that supports BIC even in a single waveguide. This type of BIC exists in a low-refractive-index waveguide on top of a high-refractive-index substrate, allowing for low-loss light guiding and routing \cite{LPR Zou}, thus benefiting hard material platforms that are not easy to fabricate, such as diamond and lithium niobate (LiNbO$_3$). Subsequently, this structure was experimentally demonstrated in polymer-on-LiNbO$_3$ hybrid waveguides and microring resonators, achieving \textit{Q} factors as high as $5.8\times10^5$ \cite{Optica XKS}. Based on this architecture, various applications have been demonstrated, such as acoustic-optic modulation, second-harmonic generation, and high-dimensional optical communication \cite{ACS XKS, LPR XKS, NC XKS}, etc.

An alternative method for generating BIC was proposed by Friedrich and Wintgen in 1985, utilizing mode coupling in the radiation channels \cite{PRA FW, PRB BIC Variable}. Notably, compared to symmetry-protected BIC, Friedrich-Wintgen BIC offers advantages in terms of device footprint and tunability. In addition to the metasurface architecture \cite{Song ACS Nano}, this type of BIC has also been investigated in individual dielectric Mie resonators by adjusting their structural parameters and demonstrated enhanced nonlinear efficiency \cite{PRL Yuri, Nanophotonics 2021, NL Yuri}. Even though the modes supported inside the Mie resonator are approximately orthogonal, their far-field profiles exhibit quite large overlap, rendering interference out of the resonator possible when their frequencies are close to degeneracy. As a result, an unexpected high-\textit{Q} factor is achieved via strong coupling of pairs of leaky modes, in which the radiation loss of one mode is dramatically suppressed with destructive interference, leading to an increased \textit{Q} factor, while the other mode with constructive interference becomes more lossy. In another widely employed optical structure - the WGM microcavity, despite the extensive study of mode coupling, there are few experimental reports on the generation of Friedrich-Wintgen BIC. Traditional WGMs differ significantly from Mie resonators as they are confined along the circular cavity boundary (i.e., microdisk and microring) through total internal reflection \cite{Nature Vahala}. Consequently, the required radiation channels for external strong coupling are naturally forbidden due to geometrical rotational symmetry. As such, a critical step toward the generation of Friedrich-Wintgen BIC in WGM microcavities is the realization of controllable radiation channels that allow WGM modes to leak and then interfere in these shared continuum.

In this study, we propose and experimentally investigate a robust and general mechanism to generate Friedrich-Wintgen quasi-BIC in a single WGM microcavity. Unlike traditional circular microcavities, we introduce boundary deformation to engineer the phase space structure, enabling shared unidirectional radiation channels through the contracted unstable manifold associated with chaotic-assisted tunneling process to facilitate strong mode coupling outside the cavity. Beyond the unidirectional emissions demonstrated in various experiments \cite{RMP Jan}, the advanced mode manipulation capabilities of deformed microcavities hold significant promise for several important applications, such as the suppression of laser spatiotemporal instabilities \cite{bittner2018suppressing}, novel approaches for generating experimental points \cite{yi2018pair}, and phase space engineering to tailor emission channels \cite{qian2021regulated}, and many others, suggesting that deformed microcavities provide greater freedom for mode control compared to traditional circular WGM microcavities. Our experimental results show avoided resonance crossing at the mode resonance frequencies, indicating strong mode coupling. Interestingly, in contrast to traditional strong coupling, the demonstrated external strong coupling exhibits a distinct characteristic in the imaginary region: the original high-\textit{Q} mode exhibits an even higher \textit{Q} factor with approximately a 3-fold enhancement, while the low-\textit{Q} mode becomes more lossy, thereby confirming the generation of Friedrich-Wintgen quasi-BIC. We present a theoretical model based on a $2\times2$ Hamiltonian matrix to analyze the formation of such Friedrich-Wintgen quasi-BIC and show excellent agreement with our experimental data. It is worth noting that the strong mode coupling demonstrated here not only modifies the real part, as routinely pursued in engineering mode dispersion, but also provides a flexible knob in manipulating the imaginary properties, resulting in, e.g., the demonstrated enhancement of the \textit{Q} factor. Our results hold significance for the fields of chaotic microcavities, nonlinear optics, quantum optics, and photonic integrated chips.

\section{Construction of Shared Leaking Continuum}

\begin{figure}[h!]
\centering\includegraphics[width=13cm]{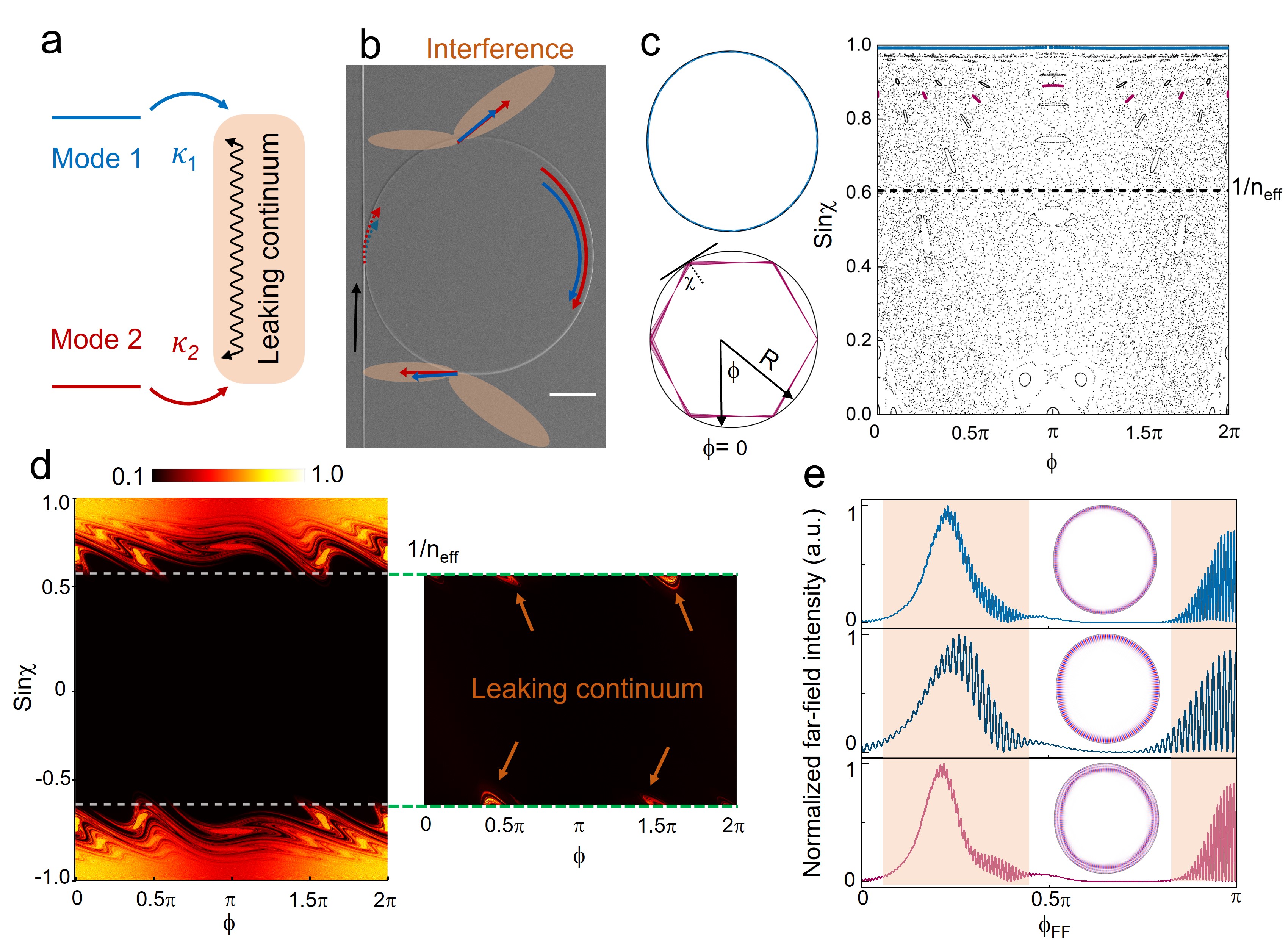}
\caption{ (a) Schematic illustration of the Friedrich-Wintgen quasi-BIC formed by external strong coupling in a shared leaking continuum. (b) SEM image of the Limaçon microdisk coupled with a bus waveguide supporting Friedrich-Wintgen quasi-BIC. The scale bar is {\SI{50}{\micro\meter}}. (c) Poincaré surface of the section of the Limaçon microcavity (without bus waveguide) at deformation parameter $\epsilon = 0.35$ and $n_{\text{eff}} = 1.649$. The side pictures show the ray trajectories of quasi-WGMs (blue) and 6-bounce (red) modes, respectively. Not all supported modes are plotted for clarity. (d) Normalized survival probability after 1500 $\times$ 2000 uniformly distributed rays bouncing 200 times inside the Limaçon microcavity, where rays with intensity above 0.1 are extracted. (e) Wave-optics simulation results of the normalized far-field spectrum of the WGM-like and 6-bounce modes. The inset pictures show the field distributions of the quasi-WGMs (top) near $kr = 106$, quasi-WGMs (middle) near $kr = 59$, and 6-bounce modes (bottom) near $kr = 106$, respectively.}

\end{figure}

The realization of Friedrich-Wintgen BIC requires a shared leaking continuum where destructive interference of radiation tails occurs, as illustrated schematically in Figure 1(a). However as mentioned earlier, this requirement is a challenge for conventional WGM microcavities due to their rotational symmetry. One possible solution would be directly attaching a channeling waveguide to the cavity, a technique known as ``end-fire injection'' with experimental validations reported in Ref.~\cite{APB Liu, Optica Liu, Light Liu}. Despite this approach's achieving high output coupling efficiencies as well as high-\textit{Q} factors, identifying a suitable parameter for detuning the system to form external strong coupling remains a daunting task. We introduce deformed microcavities to fulfill this criterion due to their nature of unidirectional emissions, a property extensively studied in both theory and experiments \cite{Nature Stone, Science Stone, PRL Jan Limacon, PRL Song 2012}.

Among the various deformed microcavity shapes, our research focuses on the Limaçon microdisk as a proof-of-principle demonstration. The cavity boundary is defined in polar coordinates by $R(\phi) = r(1+\epsilon \cos(\phi))$, where $r$ represents the cavity size and $\epsilon$ is the deformation parameter, with $\epsilon = 0$ yielding a regular circular microdisk cavity. Figure 1(b) shows a scanning electron microscope (SEM) image of the fabricated Limaçon microdisk with $\epsilon = 0.35$, utilized in subsequent numerical simulations and experimental demonstrations. Previous studies of Limaçon microcavities have demonstrated unidirectional laser emissions and their resilience to a broad spectrum of refractive indices and deformation parameters~\cite{PRL Jan Limacon}, alluding to the potential for modes to externally couple via these unidirectional emission channels and generate Friedrich-Wintgen BIC. To comprehend the behavior of light propagation inside the cavity, we project the ray trajectories onto the phase space under short-wavelength approximation, known as the Poincaré surface of section. Here, we set the effective refractive index $n_{\text{eff}}$ = 1.649 to be consistent with the subsequent experimental settings. Figure 1(c) displays the corresponding 2-dimensional phase space structures for $\epsilon = 0.35$ without the coupled bus waveguide. Clearly, this regular-chaotic mixed phase space consists of three mode families: unbroken Kolmogorov-Arnold-Moser (KAM) curves near $\sin{\chi} \approx 1$ representing the quasi-WGMs, closed curves (islands) representing the regular $p$-bounce modes, and randomly distributed dots representing the chaotic sea. From a ray optics perspective, the quasi-WGMs (green line in Fig. 1(c)) and some $p$-bounce modes (red islands for 6-bounce in Fig. 1(c)) can be confined in the cavity indefinitely, as they propagate above the total reflection angle (if $\sin{\chi} > 1/n_{\text{eff}}$). However, the quantum dynamics  allow a small probability for light rays to tunnel from the KAM or closed curves to the nearby chaotic sea and refractively escape the cavity quickly following Fresnel's laws (when $\sin{\chi} < 1/n_{\text{eff}}$). 

Although chaotic rays appear randomly distributed in the phase space, their leakage follows certain channels. In Fig. 1(d), we simulate the normalized survival probability of 1500 $\times$ 2000 rays initially uniformly distributed in the phase space structure and derive the subsequent residual intensity distributions according to Fresnel's law after 200 bounces. Major emissions occur at specific locations in the phase space structures, indicating that the supported high-$Q$ modes (quasi-WGMs and $p$-bounce modes) share the same leaking continuum. We further verify these unidirectional emission channels using wave-optics simulations based on the eigenfrequency analysis model of COMSOL. To reduce the simulation time without compromising the generality, we calculate the normalized far-field intensity spectra of two highest-\textit{Q} modes near $kr = 106$ ($k = 2\pi/\lambda$, $\lambda$ is the wavelength, and $r$ is the radius), as shown in Fig. 1(e).  Consistent with ray-optics expectations, the high-\textit{Q} modes exhibit nearly identical unidirectional emissions. Additionally, their field distributions display strong spatial overlaps, as shown in the inset figures of Fig. 1(e), where the top and middle correspond to a quasi-WGM mode and the bottom is a 6-bounce mode. It is also worth noting that the simulated quasi-WGMs and 6-bounce modes in Fig. 1(e) are representative examples and do not exactly correspond to mode A and B observed in the following experimental results. In addition, besides these two mode families, other high-$Q$ modes can also support these unidirectional emissions, as they escape following the same unstable mainfolds. Another quasi-WGM mode near $kr = 59$ is depicted in the center of Fig. 1(e), illustrating the robustness of the constructed leaky continuum across different mode types and frequency locations. Thus, by engineering the phase space structures, the shared leaking continuum that conventional circular WGM microcavity cannot support is structured, paving the way for the creation of Friedrich-Wintgen BIC.

\section{Theoretical Model for Mode Coupling}

Considering a cavity in which two modes couple with each other, this system can be represented by a $2\times2$ Hamiltonian matrix~\cite{PRL Jan ARC, RMP Jan}

\begin{equation}
\label{eq:Hamiltonian}
\begin{aligned}
H&=\begin{pmatrix}
E_1& \kappa_1\\
\kappa_2&E_2
\end{pmatrix},
\end{aligned}
\end{equation}

\noindent where $E_1$ and $E_2$ are the energies without coupling, $\kappa_1$ and $\kappa_2$ are their coupling constants. In an optical microcavity system, $E$ is a complex value with the real part $|\text{Re}\left[E_i\right]|$ referring to the resonant frequency and the imaginary part $|\text{Im}\left[E_i\right]|$ determining the loss, where $i\in\{1,2\}$ indexes the two stetes. Their $Q$ factors are expressed as $Q_i=|\text{Re}\left[E_i\right]|/2|\text{Im}\left[E_i\right]|$. Near the coupling region, the eigenenergies can be derived as

\begin{equation}
\begin{aligned}
 E_{A,B}&=\frac{E_1+E_2}{2}\pm\sqrt{\frac{\left(E_1-E_2\right)^2}{4}+\kappa_1\kappa_2}.
\end{aligned}
\end{equation}

\begin{figure}[h!]
\centering
\includegraphics[width=14cm]{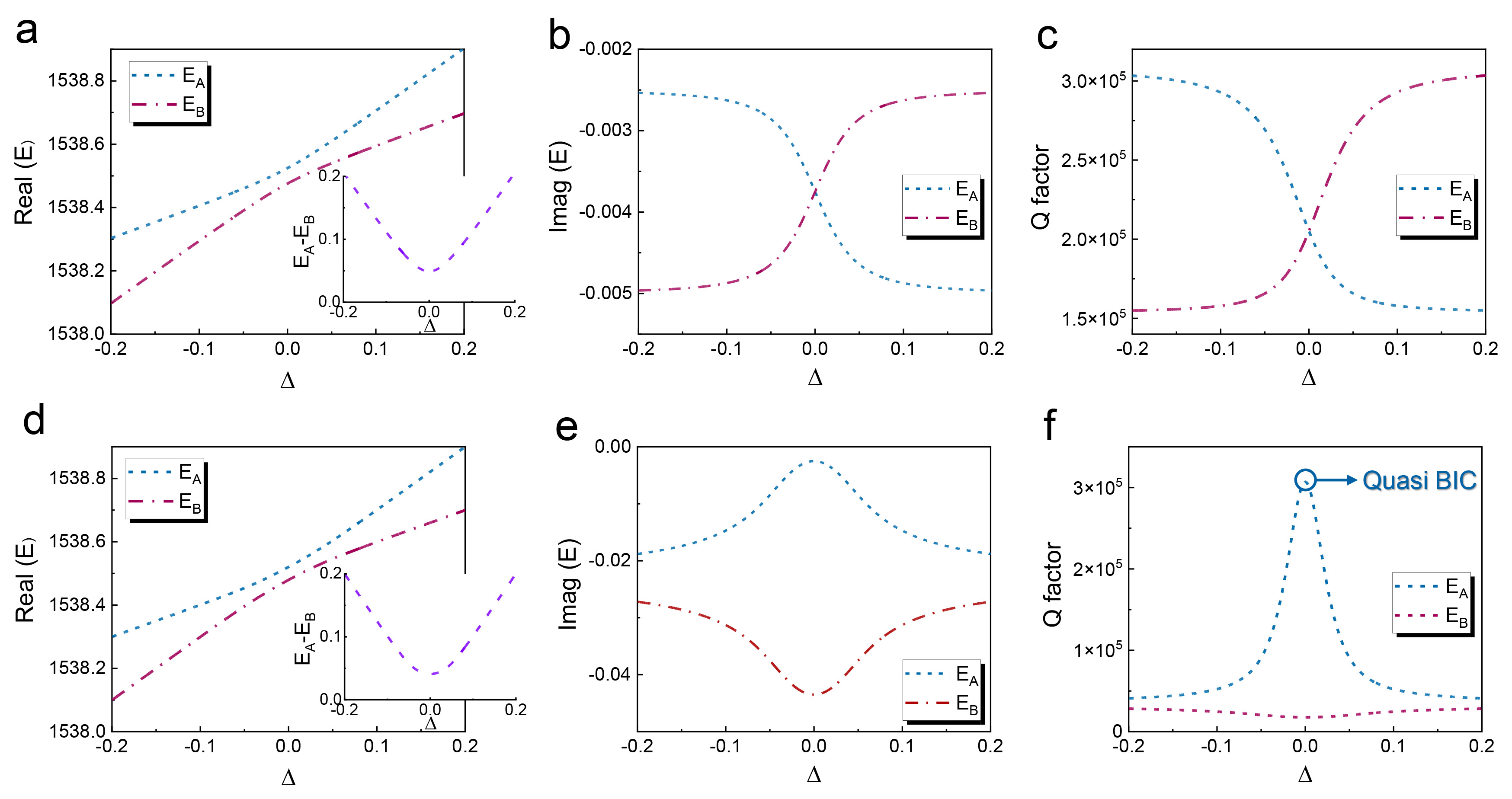}
\caption{Internal and external strong mode coupling in a microcavity. (a) and (b) are the real and imaginary parts of the eigenenergies of the Hamiltonian derived in Eq. (2) as a function of detuning  parameter $\Delta$. The inset figure in (a) shows the real eigenvalue difference. To demonstrate internal strong mode coupling, we set  $E_1 =1538.5 + \Delta -0.0025i$, $E_2 =1538.5 + 2\Delta-0.005i$, and $\kappa_1\kappa_2 = 0.0006$ as an example. (c) shows the corresponding \textit{Q} factors. (d) to (f) show the strong external coupling case, where $E_1 =1538.5 + \Delta -0.023i$, $E_2 = 1538.5 + 2\Delta-0.023i$, and $\kappa_1\kappa_2 = 0.00084i$.} 
\end{figure}

In the internal coupling situation, the states are coupled inside the cavity. The coupling constant terms are conjugates with $\kappa_1 = \kappa_2^*$ such that $\kappa_1\kappa_2$ is a positive real number. When $2\sqrt{\kappa_1\kappa_2} < |{\rm Im}(E_1)-{\rm Im}(E_2)|$, known as weak coupling, the real parts of the hybrid states will have a crossing and their imaginary parts will be an avoided resonance crossing. As the coupling strength is weak, the hybridization of the states is insignificant. Conversely, strong coupling is achieved when $2\sqrt{\kappa_1\kappa_2} > |{\rm Im}(E_1)-{\rm Im}(E_2)|$ and allows the real parts to be an avoided resonance crossing while causing the imaginary parts to cross. This strong coupling is more intriguing, as the coupling strength is sufficient to dramatically alter the properties of these states, thereby providing a powerful tool to manipulate the optical system for, e.g., engineering the mode dispersion to generate dark pulse soliton with high power efficiency~\cite{NP Xue, OL Lipson}. Figure 2 (a) and (b) illustrate the real and imaginary parts of the two eigenstates in the strong-coupling regime,  with $E_1 =1538.5 + \Delta -0.0025i$, $E_2 =1538.5 + 2\Delta-0.005i$, and $\kappa_1\kappa_2 = 0.0006$ as an example. Despite being extensively studied and utilized in WGM optical microcavities, internal strong coupling typically leads to  increase in the loss (or decrease in the $Q$ factor) as observed from the evolution of Mode B in Fig. 2(b) and (c), and many other experiments \cite{OL Lipson, OE Lipson, Andrew Optica}. Most applications, however, desire high-$Q$ modes. 

The configuration for mode coupling in a shared leaking continuum as articulated in the previous section represents a completely distinct scenario. In contrast to coupling occurring within the cavity, the two states leak into unidirectional emission channels and couple externally to the cavity via this shared continuum. For this situation, the coupling strength terms can be $\kappa_1 \neq \kappa_2^*$, and $\kappa_1\kappa_2$ is treated as a complex number. As depicted in Fig. 2(d) - (f), we illustrate an example with $E_1 =1538.5 + \Delta -0.023i$, $E_2 = 1538.5 + 2\Delta-0.023i$, and $\kappa_1\kappa_2 = 0.00084i$. The parameter $\Delta$ is introduced here to adjust the energy difference, and can be analogous to different mode numbers for various mode families or to resonance wavelength detuning induced by the thermo-optic effect as demonstrated in the following experimental findings. We also keep the total loss (coupling loss and intrinsic loss) constant to simplify the model. The real parts of the hybrid modes exhibit similar avoidance of resonance crossing as observed in the case of strong internal coupling (Fig. 2(a)). Interestingly, their imaginary parts behave differently: as the detuning parameter $\Delta$ approaches zero, where their coupling is the strongest, the initially high-$Q$ Mode A exhibits a remarkable reduction in its $\text{Im}(E)$ and an enhancement of its \textit{Q} factor, indicating a significant suppression of loss. Conversely, Mode B, with an initially low $Q$ factor, experiences a further increase in loss and a reduction in its \textit{Q} factor.  Theoretically, by carefully setting the proper coupling term, the $\text{Im}(E)$ can be close to zero, meaning the coupling to all loss channels is completely suppressed, and an infinite \textit{Q} factor can be expected. This situation is known as BIC. However, in real-world microcavity systems, not all loss channels can be completely eliminated, such as material absorption or boundary roughness-induced scattering loss. For our scenario, as marked in Fig. 2(f), the Mode A at $\Delta = 0$ indicates the formation of Friedrich-Wintgen quasi-BIC evidenced by its significant enhancement of \textit{Q} factor. 

Obviously, these clear different behaviors between internal and external strong coupling scenarios highlight the crucial role of the external leaking continuum in generating Friedrich-Wintgen quasi-BICs. The shared leakage channels, as constructed in Section 2, offer strong spatial overlap for different resonant modes, thereby providing the desired complex coupling term $\kappa_1\kappa_2$. To facilitate the formation of such Friedrich-Wintgen quasi-BIC modes, effective detuning parameters are also necessary to control the resonances and achieve strong frequency overlap, including mode numbers and temperature-induced effective refractive index modulation utilized in the following of our experimental demonstrations. It's also worthy noting that although our study focuses on the strong coupling scenario, some numerical and theoretical studies have indicated that external weak coupling also significantly influences the properties of hybrid states \cite{PRL Song 2010, PRA Song 2013}, which also differs noticeably from conventional internal weak coupling.

\section{Experimental Results}
\subsection{Device Fabrication}

We conduct the experimental demonstrations on an integrated photonic platform comprising a 335-nm stoichiometric silicon nitride (Si$_3$N$_4$) layer deposited by low-pressure chemical vapor deposition (LPCVD) on 4000-nm wet thermal silicon oxide on silicon wafers. The fabrication process begins with the initial patterning of 350-nm maN 2403 photoresist masks on the Si$_3$N$_4$ film using an e-beam lithography (EBL) system. This step is followed by appropriate developing and post-reflow processes to improve boundary roughness and dry etching selectivity. Subsequently, dry etching is performed using fluorine-based chemicals in an inductively coupled plasma reactive ion etching (ICP-RIE) system, followed by oxygen plasma treatment and hot Piranha cleaning to remove residual polymers. The SEM images are shown in Fig. 1(a). The etching depth of the Si$_3$N$_4$ layer is approximately 290 nm, in order to reduce the resonant mode numbers. Finally, the samples are manually cleaved for subsequent testing on a fiber-edge coupling setup.

\subsection{Formation of Friedrich-Wintgen Quasi-BIC}

\begin{figure}[h]
\centering\includegraphics[width=11cm]{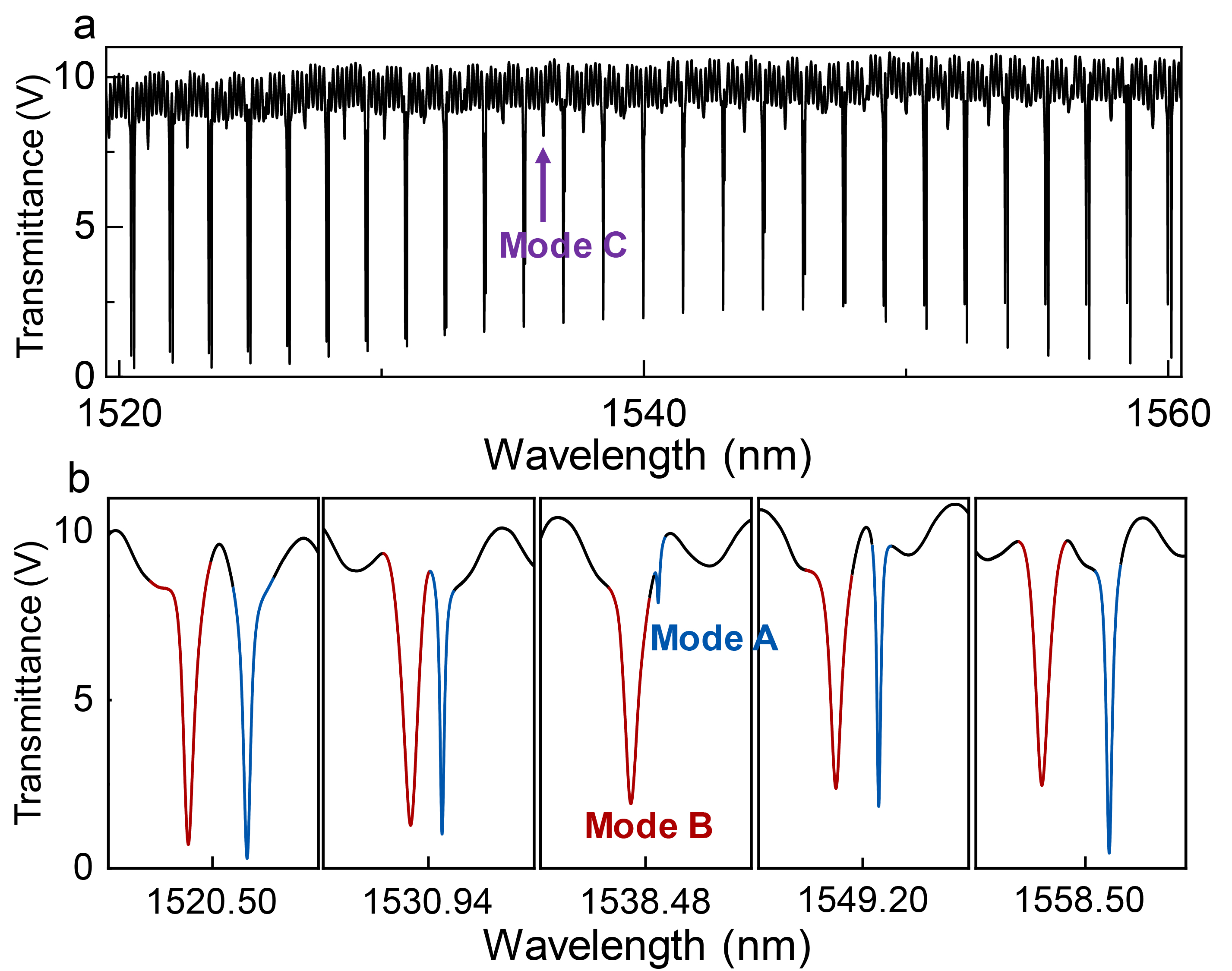}
\caption{Transmission spectrum near the avoided resonance crossing region. (a) Wide-range transmission spectrum of the Limaçon microdisk with $\epsilon = 0.35$. (b) Zoom-in view of spectra in (a) before, near, and after the strong coupling regions. Clear variations in resonant wavelength difference and resonant linewidths are observed. Each spectrum in (b) has a fixed wavelength range of 0.4 nm.}
\end{figure}

To excite the resonances, we position a bus waveguide parallel to the Limaçon microdisk at $\phi=\pi$ and adjusted the cavity size $r$, deformation parameter $\epsilon$, waveguide width $w$, and waveguide-to-cavity gap $g$ to be {\SI{120}{\micro\meter}}, 0.35, {\SI{1.5}{\micro\meter}}, and {\SI{700}{\nano\meter}}, respectively. The above coupling conditions, as well as the shallow-etched depth, are selected to balance the supported number of mode families and optical loss. This dimension corresponds to $n_{\text{eff}}$ = 1.649 for transverse electric (TE) polarization modes. Only TE polarization waveguide modes are excited by carefully adjusting a fiber polarization controller before injection. The transmission spectrum is then measured using a fiber-to-waveguide edge coupling setup.

In Fig. 3, we present the experimentally recorded transmission spectrum in the vicinity of 1540 nm, revealing the excitation of three distinct sets of modes, denoted as Mode A, B, and C. These modes belong to different mode families with varying free spectral ranges (FSRs). So, at certain region, some modes have close resonant wavelengths, thus offering strong frequency overlap to couple with each other.  Mode C is ignored in the subsequent analyses as it remains spatially isolated from Mode A and Mode B, thereby having a negligible coupling with them. From the spectrum in Fig. 3(a), it is evident that the offset between the resonances of Mode A and Mode B is clear around 1520 nm, gradually reduces as the wavelength increases, and then diverges again at longer wavelengths. Figure 3(b) is a zoom-in view on the variation in the modes' separation at different wavelengths. Notably, the offset between the two modes is correlated with their linewidths. At a larger offset around 1520.50 nm, Mode A and Mode B exhibit comparable linewidths. However, as they approach each other, the linewidth of Mode A displays a noticeable reduction near 1530.94 nm, reaching a minimum value at a wavelength of 1538.48 nm. This suggests a substantial decline in the loss Mode A perceives. After their separation, Mode A gradually broadens and recovers to its original linewidth, while Mode B exhibits a trend opposite to that of Mode A.

\begin{figure}[h!]
\centering\includegraphics[width=12cm]{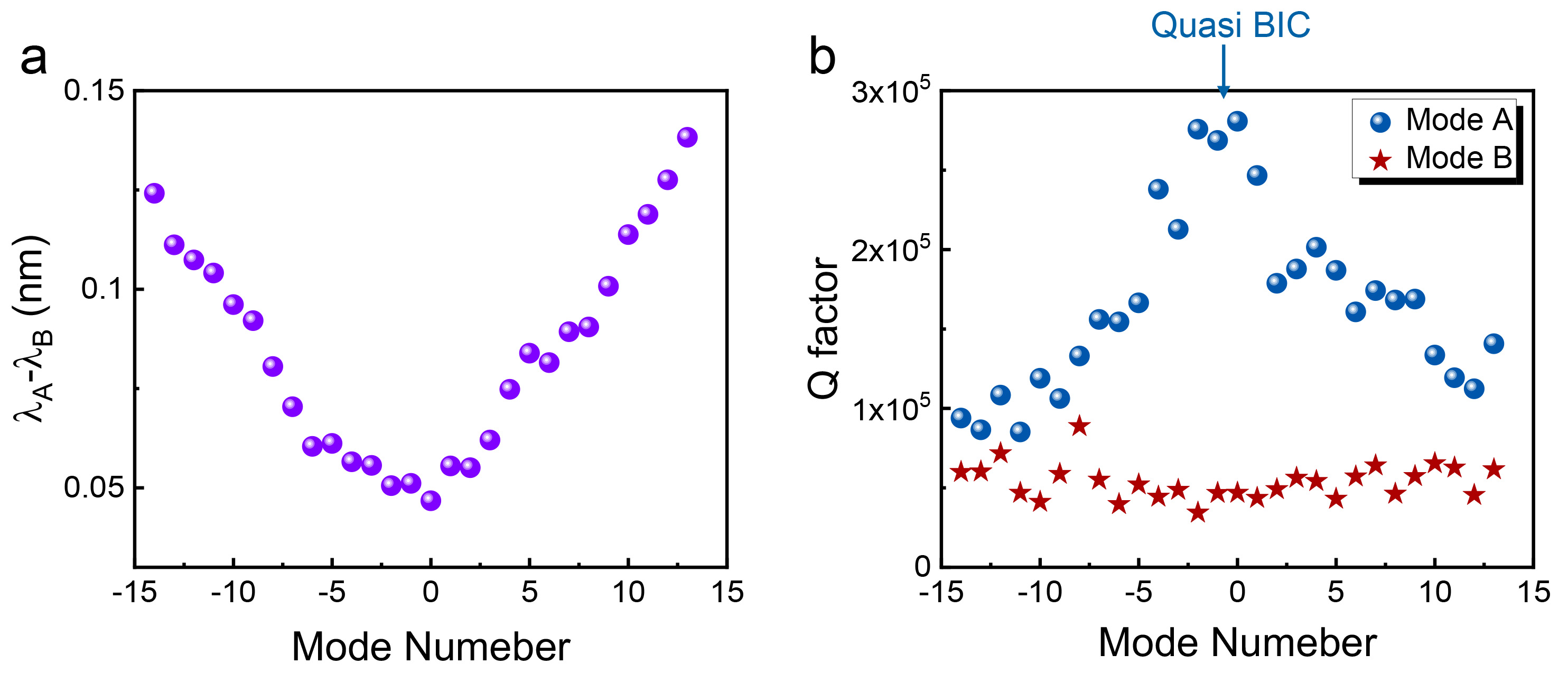}
\caption{ Formation of Friedrich-Wintgen quasi-BIC. The resonant wavelength difference in (a) and $Q$ factors of Mode A and Mode B in (b) are extracted from the spectrum in Fig. 2(a) by Lorentzian fitting. Mode number 0 corresponds to resonances at a wavelength of 1538.48 nm. Avoided resonance crossing along with the significantly enhanced \textit{Q} factor of Mode A indicates the formation of Friedrich-Wintgen quasi-BIC.
}
\end{figure}

\begin{figure}[h!]
\centering\includegraphics[width=12cm]{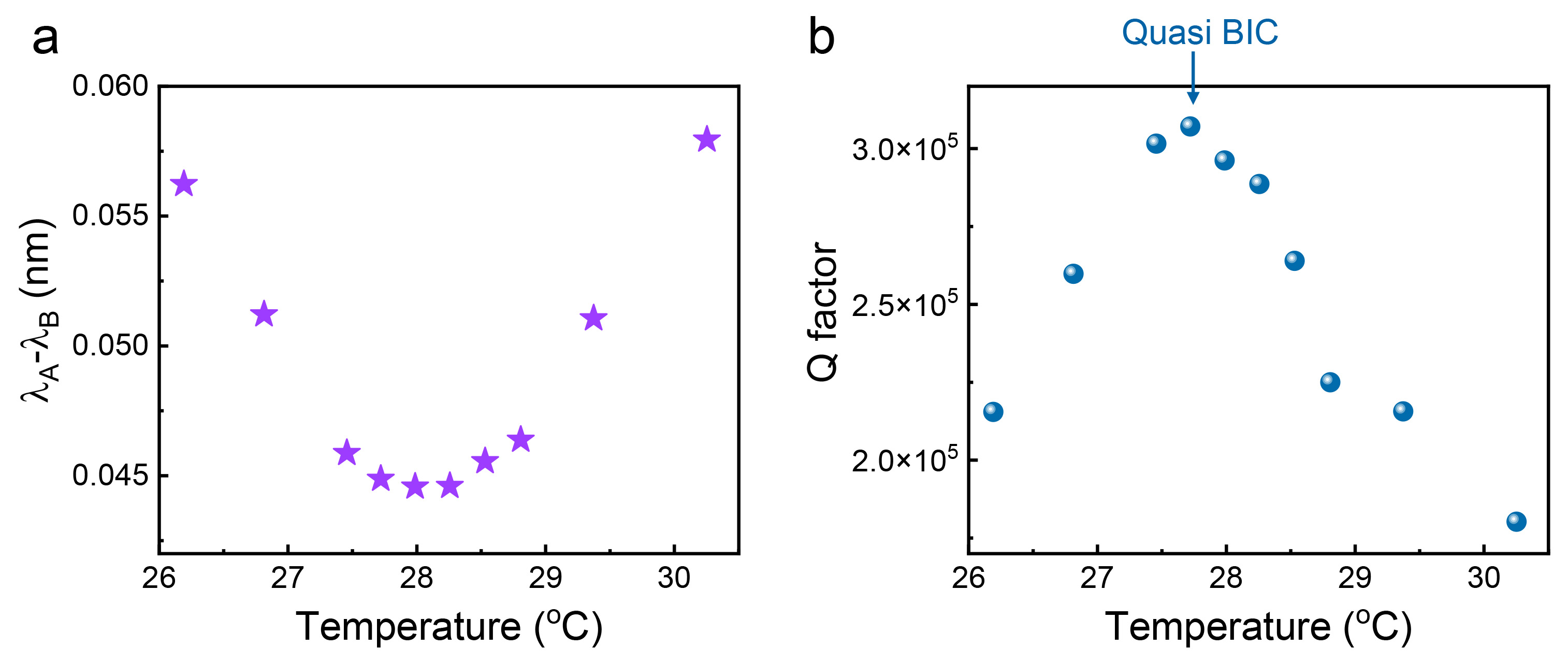}
\caption{ Fine tune of the Friedrich-Wintgen quasi-BIC at Mode number 0. (a) Resonant wavelength difference and (b) \textit{Q} factors of Mode A as a function of substrate temperature.}
\end{figure}

To grasp a more comprehensive understanding of the evolution of Mode A and Mode B in the avoided resonance crossing region, we characterized their resonant wavelength differences and individual \textit{Q} factors, corresponding to the real and imaginary parts of the eigenenergies. The resonant wavelengths clearly exhibit a pronounced avoided resonance crossing, as depicted in Figure 4(a), indicating strong mode coupling. The point where the two modes are closest, approximately at a wavelength of 1538.48 nm, is labeled as mode 0. By fitting Lorentzian curves to each resonance presented in Figure 3(a), we determine the \textit{Q} factors of Mode A and Mode B, as shown in Figure 4(b). More details about the resonance fittings are discussed in Appendix A and Fig. 7. Remarkably, with an increasing coupling strength as the two modes approach each other, we observe a significant enhancement in the \textit{Q} factor of the originally long-lived Mode A, suggesting effective suppression of the loss channel due to strong coupling in the shared leaking continuum with Mode B. Conversely, with a decreasing coupling strength as the two modes move apart, the \textit{Q} factor of Mode A is reduced while that of Mode B slightly increases. In contrast to Mode A, the \textit{Q} factors of Mode B show a slight decrease initially with increasing mode number, followed by a gradual increase. These experimental findings match well with the theory models (as shown in Fig2. (d)-(f) ) and offer a compelling evidence for the formation of Friedrich-Wintgen quasi-BIC. At this juncture, it is important to note that other loss channels, such as waveguide coupling loss, material absorption loss, and scattering loss, cannot be completely eliminated. Therefore, only Friedrich-Wintgen quasi-BIC, instead of true BIC, can be physically realized. 

In addition to the dramatically enhanced \textit{Q} factors, the decreased coupling efficiency is most likely caused by changes in their mode distributions induced by external strong coupling. This phenomenon has been also evidenced in some numerical demonstrations ~\cite{PRL Jan ARC, RMP Jan}.Visualizing the quasi-BIC modes experimentally would be highly beneficial for fully understanding the underlying physics driving the \textit{Q} factor enhancement. The recently developed direct resonance mapping technique could potentially be employed to directly measure the mode distributions near regions of external strong coupling ~\cite{Light Liu}. However, this technique is beyond the scope of our current research and not directly applicable to the Si$_3$N$_4$ platform used in this study.

 Our theory models as well as the experimental results in Fig. 3 suggest that the coupling strength affects the \textit{Q} factor of Mode A. To further explore the role of the coupling strength on Friedrich-Wintgen quasi-BIC, we utilized a thermoelectric cooler (TEC) to fine tune the refractive index of the microdisk. The quantitative experimental data in Figure 5 further corroborate the qualitative phenomena illustrated in Figure 3. By maximizing the coupling strength through TEC-induced refractive index modulation, the highest \textit{Q} factor of Mode A exceed $3 \times 10^5$, representing an enhancement of more than three times the \textit{Q} factor before external mode coupling.

\subsection{Tunability of Friedrich-Wintgen Quasi-BIC}

\begin{figure}
\centering\includegraphics[width=11cm]{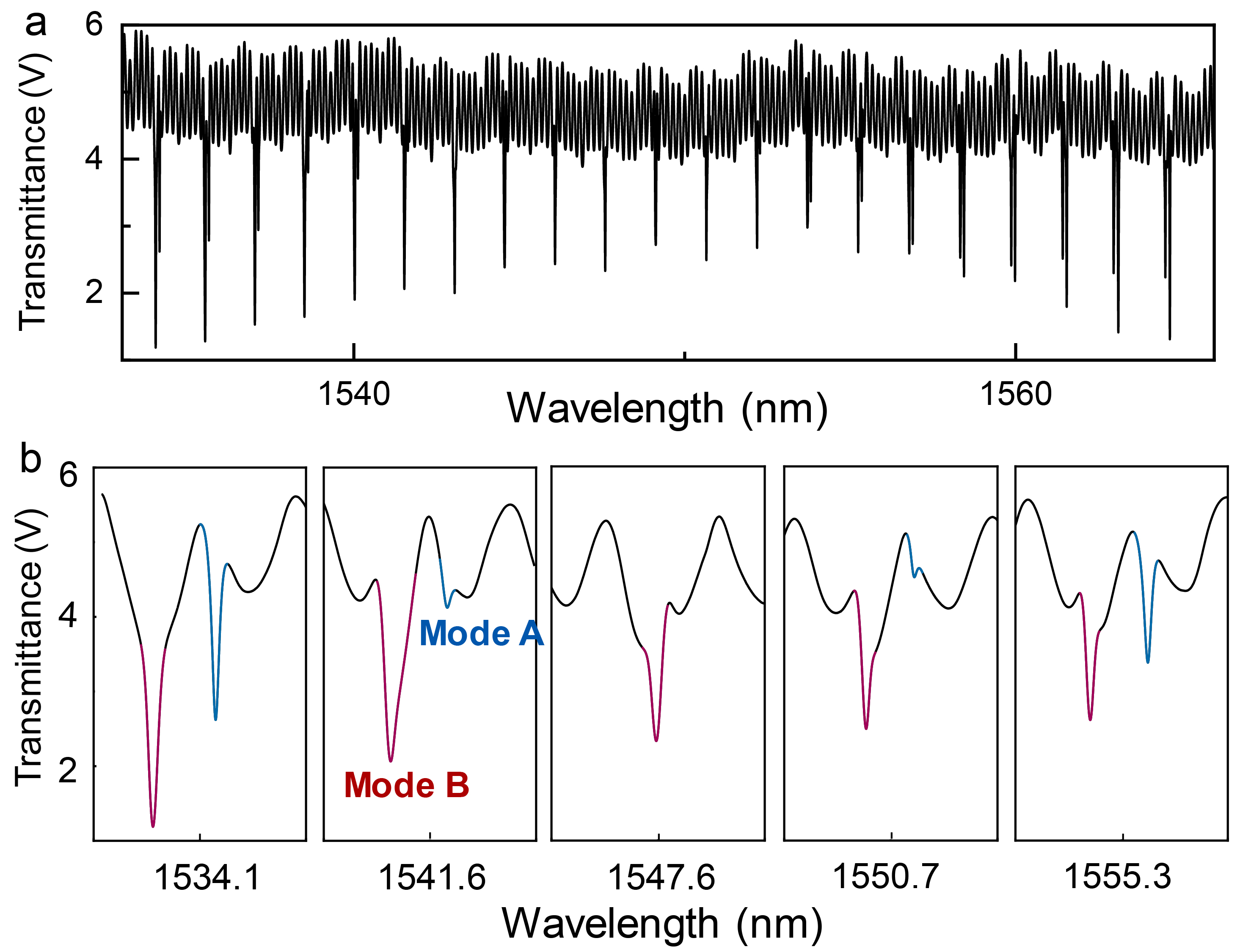}
\caption{Formation of Friedrich-Wintgen qauasi-BIC in a single Limaçon microcavity with $\epsilon = 0.4$. (a) Wide-range transmission spectrum of the Limaçon microcavity with $\epsilon = 0.4$. (b) Zoom-in view of spectra in (a) before, near, and at strong coupling. Each spectrum has a fixed wavelength range of 0.4 nm.}
\end{figure}

By constructing the unidirectional emission channels, the above experimental evidence has demonstrated the realization of Friedrich-Wintgen quasi-BIC in the Limaçon microdisk with $\epsilon = 0.35$. Intuitively, by tweaking the deformation parameter $\epsilon$, we can fine tune the phase space structures and the coupling conditions of the Limaçon microdisk, thereby engineering its emission channels and the properties of the generated Friedrich-Wintgen quasi-BIC. Previous experimental results have shown that a deformation parameter $\epsilon = 0.4$ corresponds to optimal directionality ~\cite{NJP Capasso}. To validate this hypothesis, we fabricated the Limaçon microdisk with $\epsilon=0.4$ on the same substrate, while holding all other parameters the same as those in Fig. 3. The corresponding transmission spectrum depicted in Figure 6(a) is expected to exhibit a similar avoided resonance crossing behavior to Fig. 3. However, dramatically different from the case of $\epsilon=0.35$, from wavelength of 1534.1 nm to 1547.6 nm, the resonance dip of Mode A gradually disappears as the two modes approach, indicating the presence of complete destructive interference. Subsequently, the resonance of Mode A reappears as the two modes move apart, while the coupling efficiency of Mode A also increases. This new set of data not only reaffirms the generation of Friedrich-Wintgen quasi-BIC again, but also indicates that the deformation parameter constitutes an effective means to manipulate the properties of Friedrich-Wintgen quasi-BIC.

\section{Conclusions}

In this study, we proposed and experimentally demonstrated a robust mechanism for generating Friedrich-Wintgen quasi-BIC in WGM optical microcavities. By deforming the cavity boundary, we constructed stable unidirectional emission channels shared by different resonant modes. Through this leaking continuum, we experimentally investigated external strong coupling, resulting in a significant enhancement in the \textit{Q}-factor by more than three times. While not extensively discussed in this paper, strong mode coupling also serves as a powerful tool for manipulating optical microcavity systems. It enables mode dispersion tuning, facilitates the generation of dark-pulse solitons \cite{NP Xue}, enhances power conversion efficiency in bright solitons \cite{arXiv Comb efficiency}, and improves the performance of squeezed-light sources  \cite{NC Squeezing}. Our work will create new tools for the fields of chaotic microcavities, quantum optics, nonlinear optics, and integrated photonics.

\section{Appendix A}
The measured transmission spectra reveal two types of resonant modes: those from the Limaçon microdisk and the Fabry–Pérot (FP) modes caused by reflections at the bus waveguide edges. To accurately extract the \textit{Q} factors, we apply additional FP background fitting to some resonances until achieving a high goodness of fit (GOF), quantified by an R-Square value greater than 0.99. Figure 7 summarizes some of the fitting results (resonances of Fig. 3(b)) with high GOF. All fitted \textit{Q} factors in Fig. 4(b) and 5(b) have GOF > 0.99.

\begin{figure}[h!]
\centering\includegraphics[width=10cm]{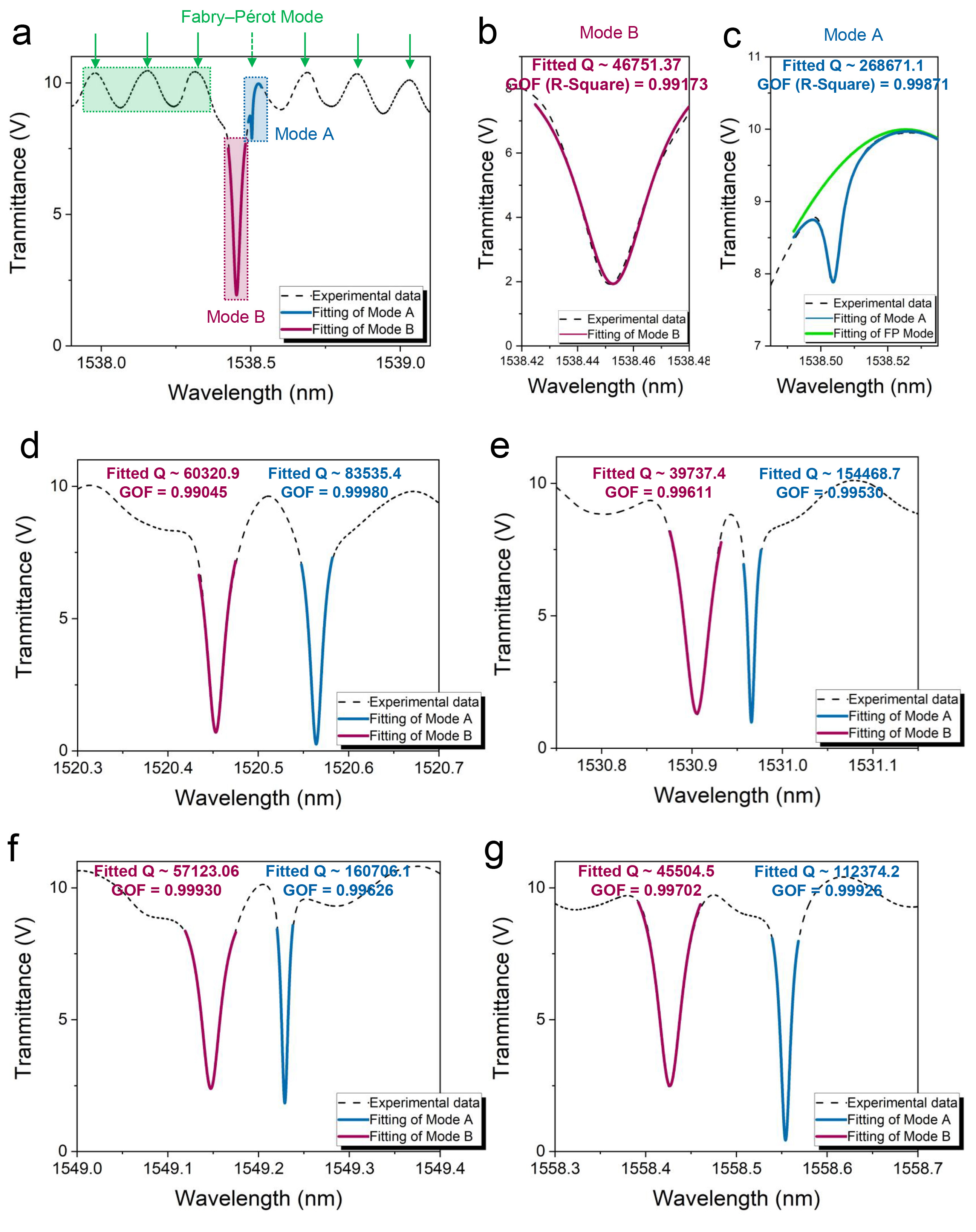}
\caption{  Resonance fitting and \textit{Q} factor extraction: (a) Transmission spectra near the quasi-BIC region, with green arrows indicating the Fabry–Pérot background caused by reflections at the bus waveguide edges. (b) and (c) Fitting results for Mode B and Mode A, respectively. (d) - (g) Fitting results for additional resonances.}
\end{figure}

\section{Funding}
We gratefully acknowledge funding support from the National Science Foundation Grant No.~2326780 and No.~2317471 and the University of Michigan.

\section{Acknowledgments}
We thank Siyuan Zhang for helping with the C++ coding of the ray-optics simulation.

\section{Disclosures}
The authors declare no conflicts of interest.

\section{Data Availability}
All the data used in this study are available from the corresponding author upon reasonable request.


\end{document}